\documentclass[conference]{IEEEtran}
\IEEEoverridecommandlockouts
\usepackage{cite}
\usepackage{amsmath,amssymb,amsfonts}
\usepackage{algorithmic}
\usepackage{graphicx}
\usepackage{textcomp}
\usepackage{pgfplots}
\usepackage{booktabs}
\usepackage{balance}
\usepackage{multicol}
\usepackage{float}
\usepackage{color, colortbl} % allow highlighting rows in tables

\def\BibTeX{{\rm B\kern-.05em{\sc i\kern-.025em b}\kern-.08em
    T\kern-.1667em\lower.7ex\hbox{E}\kern-.125emX}}
    
\newcommand{\solution}{\textit{RansomAI}}

\newcommand{\eg}{\textit{e.g., }}  % from the Latin exempli gratia = for example
\definecolor{LightGray}{gray}{0.85}

\def \1{\textit{(i)}}
\def \2{\textit{(ii)}}
\def \3{\textit{(iii)}}
\def \4{\textit{(iv)}}
\def \5{\textit{(v)}}

\begin{document}
\author{
    \IEEEauthorblockN{Jan von der Assen\IEEEauthorrefmark{1}, Alberto Huertas Celdrán\IEEEauthorrefmark{1}, Janik Luechinger\IEEEauthorrefmark{1}\\  Pedro Miguel Sánchez Sánchez\IEEEauthorrefmark{2}, G\'er\^ome Bovet\IEEEauthorrefmark{3}, Gregorio Mart\'inez P\'erez\IEEEauthorrefmark{2}, Burkhard Stiller\IEEEauthorrefmark{1}}
    
    \IEEEauthorblockA{\IEEEauthorrefmark{1}Communication Systems Group CSG, Department of Informatics, University of Zurich UZH, CH--8050 Zürich, Switzerland \\{[vonderassen, huertas, stiller]}@ifi.uzh.ch, janik.luechinger@uzh.ch}
    \IEEEauthorblockA{\IEEEauthorrefmark{2}Department of Information and Communications Engineering, University of Murcia, 30100--Murcia, Spain {gregorio}@um.es}
    \IEEEauthorblockA{\IEEEauthorrefmark{3}Cyber-Defence Campus, armasuisse Science \& Technology, CH--3602 Thun, Switzerland gerome.bovet@armasuisse.ch}
}
% change symbols of affiliations to numbers
\DeclareRobustCommand*{\IEEEauthorrefmark}[1]{%
  \raisebox{0pt}[0pt][0pt]{\textsuperscript{\footnotesize #1}}%
}

\title{\solution{}: AI-powered Ransomware\\ for Stealthy Encryption}

\maketitle

\begin{abstract}
Cybersecurity solutions have shown promising performance when detecting ransomware samples that use fixed algorithms and encryption rates. However, due to the current explosion of Artificial Intelligence (AI), sooner than later, ransomware (and malware in general) will incorporate AI techniques to intelligently and dynamically adapt its encryption behavior to be undetected. It might result in ineffective and obsolete cybersecurity solutions, but the literature lacks AI-powered ransomware to verify it. Thus, this work proposes \solution{}, a Reinforcement Learning-based framework that can be integrated into existing ransomware samples to adapt their encryption behavior and stay stealthy while encrypting files. \solution{} presents an agent that learns the best encryption algorithm, rate, and duration that minimizes its detection (using a reward mechanism and a fingerprinting intelligent detection system) while maximizing its damage function. The proposed framework was validated in a ransomware, Ransomware-PoC, that infected a Raspberry Pi 4, acting as a crowdsensor. A pool of experiments with Deep Q-Learning and Isolation Forest (deployed on the agent and detection system, respectively) has demonstrated that \solution{} evades the detection of Ransomware-PoC affecting the Raspberry Pi 4 in a few minutes with $>$90\% accuracy.
\end{abstract}

\begin{IEEEkeywords}
Ransomware, Reinforcement Learning, Artificial Intelligence, Malware, Evasion
\end{IEEEkeywords}

\section{Introduction}
\label{intro}

With the growing progress of digitalization, companies have become more dependent on information systems that uphold their business missions. It has influenced a recent increment in malware-based attacks targeting heterogeneous enterprises~\cite{wef}. Among existing attack vectors, ransomware is one of the most significant threats affecting companies due to its impact on data and economic losses. In 2022, industries experienced a rise of 87\% in ransomware attacks~\cite{bnn}. Moreover, although ransom payments are slowly declining~\cite{bnn}, ransomware is still a highly impactful threat to most companies. As an example, IBM assessed in 2022 that the average loss of a ransomware attack was 4.54M USD~\cite{ibm}.

Detecting cyberattacks and ransomware, in particular, is the first step toward mitigating their impact. As outlined by recent work~\cite{survey}, dynamic detection approaches incorporating behavioral data into machine and deep learning (ML and DL) techniques have demonstrated to be highly effective against ransomware. However, while these behavioral-based approaches are not vulnerable to obfuscation (as static approaches), they rely on specific assumptions. In particular, both classification and anomaly detection approaches assume that malicious behavior is stable and static enough to allow ML/DL models to differentiate it from normal or benign behavior~\cite{specforce}. 

However, integrating Artificial Intelligence (AI) into ransomware samples could change it by adding intelligent dynamicity to encryption behaviors. It would complicate the detection since ransomware samples could learn the encryption rate that maximizes impact while minimizing its detection. Furthermore, the encryption rate could automatically change depending on the device status. Therefore, the literature presents an open challenge that focuses on studying the suitability of integrating AI-based techniques into ransomware to learn how and when to encrypt real devices while staying undetected. This challenge is vital to later, in a second stage, evaluate the detection capabilities of current cybersecurity mechanisms and update them if needed.

This work fills this research gap and proposes \solution{}, a framework using Reinforcement Learning (RL) and fingerprinting to maximize the stealth capabilities of ransomware samples while maximizing their damage function. \solution{} contains an RL-based agent that learns the optimum ransomware behavior (encryption rate, duration, and algorithm combination) according to the reward received for each ransomware configuration and the device behavior. The reward is calculated according to the output of a fingerprinting and ML-based anomaly detection system and the encryption rate of the selected ransomware configuration. A prototype of \solution{} (available in \cite{ransomAI}) that implements Deep Q-Learning and Isolation Forest (for the agent and the anomaly detector, respectively) has been integrated into a real ransomware called Ransomware-PoC. The prototype effectiveness has been evaluated in a Raspberry Pi 4, acting as a crowdsensor through a pool of experiments with six configurations of Ransomware-PoC dealing with the encryption algorithm, rate, and duration. The experiments have demonstrated that \solution{} evades the detection of Ransomware-PoC in a few minutes and with high accuracy.

The remainder of this paper is structured as follows. Section~\ref{sec:related} introduces the background and related approaches. Then Section~\ref{sec:problemstatement} presents the problem and scenario tackled by Section~\ref{sec:solution} with the framework architecture and its implementation. The performance of the solution is demonstrated in the experiments presented in Section~\ref{sec:experiments}. Finally, Section~\ref{sec:conclusions} draws conclusions from the results.
\section{Related Work}
\label{sec:related}

\begin{table}[b]
    \caption{Related Work Using AI for Offensive Purposes}
    \begin{tabular}{@{}lllllllll@{}}
        \toprule
        \textit{Paper}  & \textit{Attack} & \textit{Obf.} & \textit{Evasion} & \textit{Tech.} & \textit{Execution} & \textit{Eval.} \\\midrule
%        \cite{pan_aiInMalware}   & & & & & & \\
%        \cite{thanh_surveyOnAIInMalware}   & & & & & \\
%        \cite{kaloudi_surveyAICyberThreats}   & & & & & & \\
%        \midrule
        \cite{anderson_evadeStaticPEMLMalwareModels} 2018   & Adv. & Yes & Static & RL & Offline & Sim. \\
        \cite{song_mabMalware} 2022  & Adv. & Yes & Static & RL & Offline & Real \\
        \cite{castro_aimedRL} 2021 & Adv. & Yes & Static & RL & Offline & Sim. \\
        \cite{castro_aimed} 2019 & Malw. & Yes & Static & GA & Offline & Sim. \\
        \cite{chung_smartMalware} 2019  & Malw. & No & - & ML & Online & Simul. \\
        \cite{jha_mlDrivenMalware} 2020 & Adv. & No & - & ML & Online & Sim. \\
        \cite{stoecklin_deeplocker} 2022  & Malw. & Yes & Hybrid & DL & Offline & Real \\
        
        This & Malw. & No & Dynamic & RL & Online & Real \\
        \bottomrule
    \end{tabular}
    \label{tab:summaryRelatedWork}
\end{table}

In contrast to many works that demonstrate the applicability of AI for defense, there are only a few papers discussing the offensive perspective. \tablename~\ref{tab:summaryRelatedWork} compares the most relevant and recent ones. In \cite{anderson_evadeStaticPEMLMalwareModels, song_mabMalware, castro_aimedRL}, the authors leveraged RL to evade a static detection system. Here, static detection refers to an ML-based system that detects whether a piece of software is malicious based on the static structure of the malware sample. Thus, the three approaches consider RL to find an optimal way to structure the malware sample before executing the payload on the victim's premises. This is achieved by relying on an adversarial technique to evade the ML-based detection system. None of these solutions consider ransomware, and only~\cite{song_mabMalware} presents experimental results in a realistic scenario where commercial antivirus systems were deployed in cloud-based virtual machines.

Aside from using RL, \cite{castro_aimed} presented how Genetic Algorithms can help with byte-level modifications to evade malware detection. \cite{chung_smartMalware} proposed an ML model to inject strategic system failure into a cyber-physical system. The model was not used to address evasion but to optimize the time and location of the failure. Therefore, no adversarial techniques were used in that work. \textit{RoboTack} was proposed in \cite{jha_mlDrivenMalware} to estimate attacks impact instead of evading detection. A neural network with three hidden layers gave insight into the optimal deployment target, time, and strategy. Like the previous two approaches, \textit{RoboTack} was only evaluated in a simulated environment.

While all aforementioned evasion approaches deal with a static detection system, \textit{DeepLocker} \cite{stoecklin_deeplocker} is the first generic malware that does not rely on static obfuscation for evasion. In contrast, it achieves evasion by employing dynamic obfuscation of arbitrary sources before the breach event. \textit{DeepLocker} encrypts the payload and injects itself into the target system, making static analysis infeasible. The trigger conditions are transformed into a deep neural network (DNN), encoding it in another black box system. The DNN was trained on multiple target markers (\eg system-level features, geolocation, input and output systems). Upon recognizing the target, these markers can be converted into the encryption key, unleashing the payload. \textit{DeepLocker} was evaluated in a real scenario.

In conclusion, while several approaches leverage AI for malware, most assume a static defense model. Furthermore, most solutions rely on adversarial attacks for evasion, thereby focusing on optimizing the malware before its execution. Finally, most works do not present evaluation in realistic scenarios, and no work is focused on ransomware, which is currently one of the most harmful malware families. %This work aims to overcome this by evading a dynamic, behavior-based black-box detection system without relying on adversarial attacks.

\section{Problem Statement \& Scenario Description}
\label{sec:problemstatement}

Some of the most recent and promising cybersecurity solutions combine fingerprinting and ML to successfully detect anomalies and classify ransomware samples~\cite{huertas:2022:ransomware}. However, they do not consider ransomware adapting its encryption mechanisms according to specific criteria to stay undetected. Therefore, this work seeks to answer the following question: Is it possible to effectively and automatically evade novel detection systems by incorporating AI-based techniques into ransomware? Here, the main goal of AI would be to learn and select in an online and autonomous fashion what malicious configuration maximizes encryption while minimizing its detection. In addition, if evasion is possible, the next question would be: How much time is needed to adapt the ransomware behavior? This work considers the following scenario to tackle these questions.

\begin{itemize}
    \item A Raspberry Pi 4 acting as a real sensor of ElectroSense~\cite{electrosense}, a platform that monitors the radio frequency spectrum. The Raspberry Pi would host the intelligent fingerprinting detection system that must be evaded. More information about the potential detection mechanism can be found in~\cite{huertas:2022:ransomware}.
    
    \item An extended version of Ransomware-PoC composed of i) a client running on the Raspberry Pi and encrypting files by using different algorithms, rates, and duration; and ii) a command and control (C\&C) server that selects between six different configurations (see \tablename~\ref{tab:ransomwareConf}) the encryption setup executed by the client. It is important to mention that the C\&C only has control over the client, so it is not able to change the configuration of the Raspberry Pi or the anomaly detector.
    
\end{itemize}

\begin{table}[b]
    \caption{Ransomware Configurations}
    \begin{tabular}{c | c c c c }
    \toprule
        \textit{Conf.} & \textit{Algorithm} & \textit{Rate (B/s)} & \textit{Burst Duration} & \textit{Burst Pause (s)} \\\midrule
        1 & ChaCha20 & 16 & 1 file & 60 \\
        2 & AES-CTR & 565'565.65 & unlimited & 0 \\
        3 & Salsa20 & 632'834.80 & unlimited & 0 \\
        4 & AES-CTR & 500 & 10 s & 5 \\
        5 & ChaCha20 & 200 & 20 s & 40 \\
        6 & Salsa20 & 200 & 120 s & 30 \\
        \toprule
    \end{tabular}
    \label{tab:ransomwareConf}
\end{table}

\section{\solution{} Architecture}
\label{sec:solution}

\begin{figure}[t]
    \centering
    \includegraphics[width=0.7\linewidth]{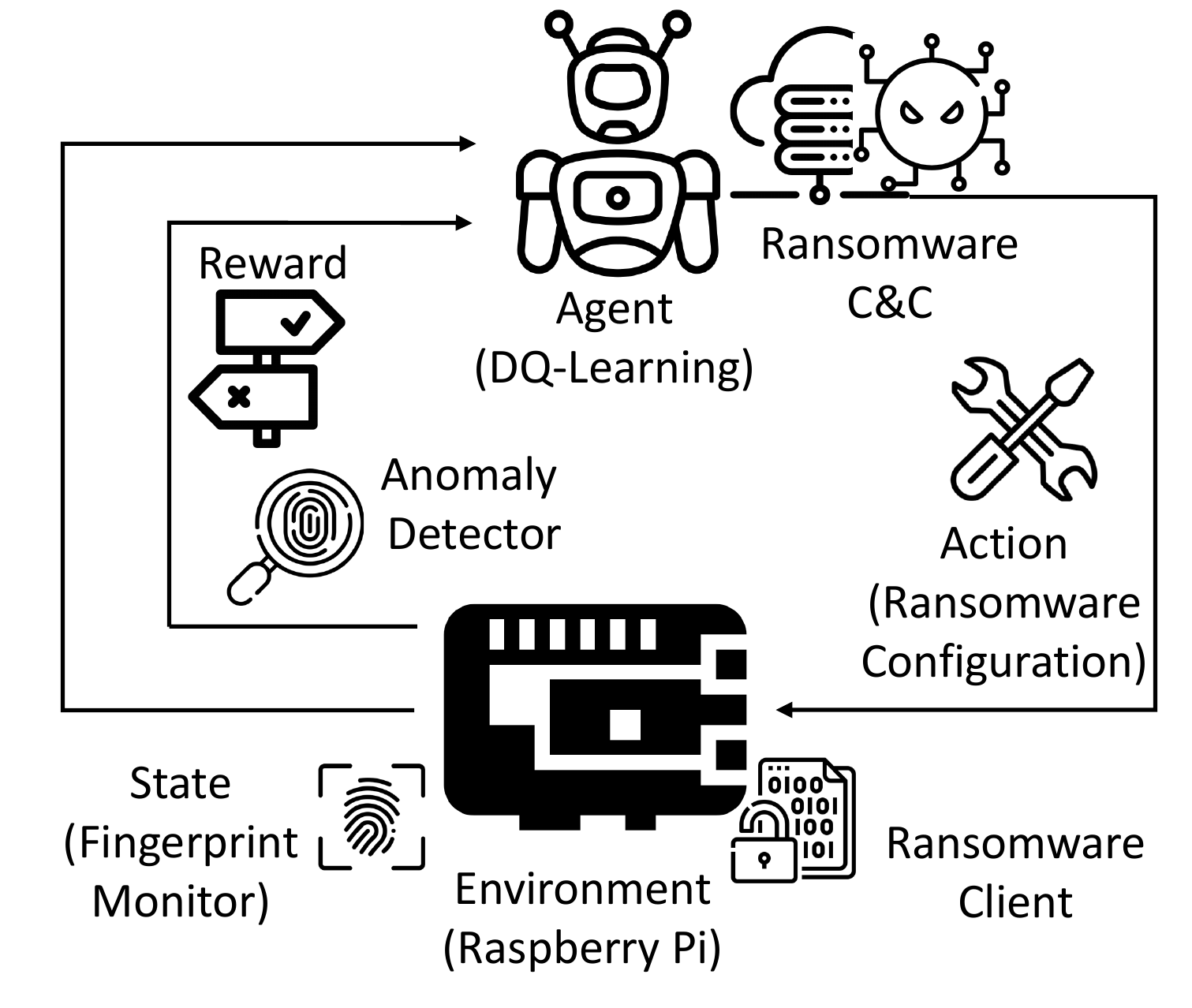}
    \caption{\solution{} Architecture}
    \label{fig:architecture}
\end{figure}

This section presents the design and implementation details of \solution{}, a framework adding intelligence to existing ransomware samples to evade detection systems while maximizing encryption. \figurename~\ref{fig:architecture} shows the main components of \solution{}, which is available in \cite{ransomAI}. 

In summary, the Agent uses RL to learn and deploy (interacting with the C\&C) the best ransomware configuration (from a list of existing ones, see \tablename~\ref{tab:ransomwareConf}). The learning process is driven by the feedback provided by a Reward function, which prioritizes the ransomware configuration that maximizes the encryption rate while minimizing its detection. For that, the Reward function computes the output of an Anomaly Detector that uses behavioral fingerprinting and ML, and the encryption rate of the selected ransomware configuration. More in detail, the Anomaly Detector (which tries to mimic a genuine system potentially deployed in the Raspberry Pi) compares the current state (behavior) of the Raspberry Pi with its normal one (monitored when it is not under attack). The Fingerprint Monitor is responsible for continuously gathering the Raspberry Pi states. All the previous components are deployed on the server where the C\&C runs, except the Fingerprint Monitor, which runs on the Raspberry Pi with the Ransomware client. More details about each component are provided below.

\subsection{State}

A state is the agent vision of the environment (Raspberry Pi) at a given time. Modeling states precisely is crucial to allow the agent to understand the impact of each action on the environment and learn proper actions. 

\solution{} proposes behavioral fingerprinting to represent states. In particular, it uses the \textit{perf} Linux command to collect different events from \textit{system calls}, \textit{CPU}, \textit{device drivers}, \textit{scheduler}, \textit{network}, \textit{file system}, \textit{virtual memory}, and \textit{random numbers} families. The reason for selecting these heterogeneous data sources is to detect small perturbations produced by encryption phases and later evade robust detection systems. More in detail, 50 features were chosen from the previous families. For this selection, 103 features were initially monitored in time windows of 5 s (decided according to previous work~\cite{huertas:2022:ransomware}) during 8 hours of Raspberry Pi normal behavior (sensor without being attacked). Then, duplicated, temporal, and constant features were removed during a data cleaning process. All remaining features were plotted, and 28 features whose normal behavior was volatile overall fingerprints were manually identified and subsequently dropped. Finally, features with more than 99\% correlation were removed. After the whole process and as mentioned, 50 features were selected to create the device fingerprints.

\subsection{Action}

Actions are the way in which the agent interacts with the environment. In \solution{}, actions correspond to the execution of the six ransomware configurations created in the extended version of Ransomware-PoC (see \tablename~\ref{tab:ransomwareConf}). It is important to mention that the original version of Ransomware-PoC provides a fixed encryption configuration (in terms of algorithm and rate), and this work extended it with new functionality in terms of different algorithms, encryption rates, and pauses between bursts.

\subsection{Anomaly Detector}

The Anomaly Detector component is critical for \solution{} since it tries to mimic the functionality of existing novel detection systems that might be deployed on the Raspberry Pi to detect ransomware attacks. Specifically, its outputs are used by the Reward function to provide positive or negative feedback for each encryption configuration. 

\solution{} proposes the combination of unsupervised ML and behavioral fingerprint. More in detail, it considers an anomaly detection system to model the normal behavior of the Raspberry Pi (acting as a sensor of ElectroSense) and detect anomalies produced by each ransomware configuration. The behavior of the device is represented by the 50 features selected to model the environment state (previously described). These features cover as many different data sources as possible to detect changes in device resource usage. Then, different unsupervised ML models such as Isolation Forest, Autoencoder, Local Outlier Factor, and One Class-Support Vector Machine were trained with the normal behavior of the device (running for 8 hours) and evaluated with the normal and six ransomware configurations. Isolation Forest was selected for the prototype implementation after analyzing the performance of each model when detecting normal and under-attack behaviors. More in detail and as can be seen in \tablename~\ref{tab:IFperformance}, normal behavior and configurations 2 and 3 were correctly detected with almost 89\% True Negative Rate (TNR) and 0\% False Negative Rate (FNR), respectively. Then, configurations 1, 4, 5, and 6 were incorrectly detected as normal behavior with 77-91\% FNR. These results were achieved with a 5\% contamination factor in Isolation Forest hyperparameters. At this point, it can be concluded that it is feasible to encrypt files without being detected by novel works. However, evaluating if \solution{} can automatically learn the optimum encryption configuration within an acceptable time is still essential.

\begin{table}[b]
    \centering
    \caption{TPR and FNR of Isolation Forest}
    \begin{tabular}{c c c c c c c }
        \toprule
        \textit{Normal} & \textit{Conf. 1} & \textit{Conf. 2} & \textit{Conf. 3} & \textit{Conf. 4} & \textit{Conf. 5} & \textit{Conf. 6} \\\midrule  
        88.94 & 91.62 & 0.62 & 0.21 & 80.18 & 82.05 & 77.38 \\   
        \toprule
    \end{tabular}
    \label{tab:IFperformance}
\end{table}

\subsection{Reward}

Positive and negative rewards allow the Agent to learn if selected ransomware configurations are good or bad. \solution{} provides rewards according to the output of the anomaly detector and the encryption rate of the selected configuration. In other words, if one ransomware configuration is not detected, a higher encryption rate should be more favorable than a lower encryption rate. Moreover, if the ransomware configuration is detected, a lower encryption rate is worse than a higher one. In summary, encryption time is important and considered by the reward mechanism.

\solution{} proposes two separate reward functions, one for hidden and another for detected encryption. After several experiments and fine-tuning, $R_{hid}(r) = 10 * ln(r+1) + h$ and $R_{det}(r) = \frac{-d}{max(r, 1)} - d$ are the two proposed reward functions. In both functions, $h$ and $d$ are constants aiming to distinguish clearly between rewards for hidden and detected behaviors, and $r$ is the current encryption rate. The constants were set as $h = 0$ and $d = 20$ to avoid impacting the weights in the network with unnecessarily high rewards. Therefore, with the reward functions and the probability of being detected (see \tablename~\ref{tab:IFperformance}), the Agent should learn that configuration 4 is the most convenient because it achieves the best expected average reward ($0.8018 * 62.166 + 0.1982 * (-20.04) = 45.87$). More in detail, following configuration 4, Ransomware-PoC could encrypt approximately 200 KB per minute, 12 MB per hour, 288 MB per day, and 8.6 GB per month without being detected.

\subsection{Agent}

%Background about RL
The agent of \solution{} considers RL and learns following a trial-and-error approach. When the agent takes one action (selects a ransomware configuration), the environment (Raspberry Pi) changes to a new state (behavioral fingerprint called afterstate) and receives a reward. This loop is repeated, and sequences of the previous steps (state, action, afterstate, and reward) are called episodes. An episode concludes when no more actions can be taken. 

In \solution{}, since the state space is vast due to the number of features (50) modeling the behavior fingerprint and their continuous values, a tabular approach for the RL model is not feasible. Instead, a neural network is used together with the Deep Q-Learning algorithm. Q-learning is an off-policy temporal difference control algorithm defined by $Q(S_t,A_t) \leftarrow Q(S_t,A_t) + \alpha \big[R_{t+1} + \gamma\ max_a\ Q(S_{t+1},a) - Q(S_t,A_t)\big]$. $Q$ is the learned action-value function that approximates the optimal action-value function based on value estimates of state-action pairs $(S,A)$ and reward $R$ independent of the followed policy. In addition, the agent follows an epsilon-greedy policy that decides with probability $\epsilon$ to take a random action, which may not coincide with the current estimated optimal action. This policy ensures the agent explores different actions to find the optimum one. Regarding the Agent neural network, it has three fully-connected layers. The input layer has 50 neurons (matching the features of the fingerprint), the hidden layer has 25 neurons, and the output layer size is 6, corresponding to the number of ransomware configurations. The details regarding the hyperparameters selection, such as exploration versus exploitation ($\epsilon$ and $\delta$), learning rate ($\alpha$), discount factor ($\gamma$), hidden neurons, activation function, and weights initialization, are provided in Section~\ref{sec:experiments}.
\section{Experiments}
\label{sec:experiments}

This section evaluates the performance of \solution{} while learning the optimum configuration to encrypt the Raspberry Pi presented in Section~\ref{sec:problemstatement}. For that, the First experiment performs an individual search of hyperparameters to optimize the agent accuracy and learning time. Then, the second shows the learning accuracy across time.

\subsection{Hyperparameters Search}

The goal of this experiment is to find the optimum configuration of hyperparameters to maximize accuracy and reduce the training time of \solution{}. In this sense, the following hyperparameters of the RL-based Agent are considered: i) exploration settings ($\epsilon$, $\delta$, $\alpha$, and $\gamma$), ii) number of hidden neurons in the neural network, iii) activation functions used in the forward and backward pass of the neural network, and iv) method for weight initialization in the neural network. The following baseline configuration is fixed while one hyperparameter per test is fine-tuned. Episodes = 5,000, $\epsilon$ = 0.4, $\delta$ = 0.01, $\alpha$ = 0.0050, $\gamma$ = 0.10, hidden neurons = 40, activation = \textit{Log and SiLU}, weights = \textit{He}.

\subsubsection{\textbf{Exploration vs. Exploitation}} It presents the trade-off between exploring new actions to discover new states and exploiting learned optimal actions. For different exploration and exploitation configurations, \tablename~\ref{tab:exploration} shows the average learning time of the Agent, the accuracy after training, the number of episodes required to find the best encryption configuration, and the average Q-value difference (AQD). As can be seen, the exploration does not significantly impact the training or learning duration, as most values for the average training time are similar. Instead, the impact is manifested in the final accuracy as well as the number of episodes and AQD. The best setting $(\epsilon=0.20, \delta=0.01)$ has the highest performance considering the trade-off for exploration and exploitation. It offers great accuracy, 99.63\%, close to the maximum of 99.84\%, needs one of the lowest numbers of episodes, and achieves high differences between the Q-values of configuration 4 and the closest ones (5 and 6).

\begin{table}[htpb!]
    \centering
    \caption{Agent Performance for Different Exploration Settings}
    \begin{tabular}{l | c c c c }
        \toprule
        \textit{Settings} & \textit{Learning (min)} & \textit{Accuracy} & \textit{Episodes} & \textit{AQD}\\\midrule
        $\epsilon=0.10, \delta=0.01$ & 140.2 & 99.84\% & 50 & 80.1\\
        \rowcolor{LightGray}$\epsilon=0.20, \delta=0.01$ & 141.8 & 99.63\% & 40 & 125.1 \\
        $\epsilon=0.30, \delta=0.01$ & 139.1 & 99.34\% & 40 & 60.5\\
        $\epsilon=0.40, \delta=0.01$ & 139.4 & 99.24\% & 150 & 60.5 \\
        $\epsilon=0.50, \delta=0.01$ & 126.7 & 99.24\% & 300 & 59.5\\
        $\epsilon=0.10, \delta=0.001$ & 127.0 & 98.75\% & 400 & 33.3\\
        $\epsilon=0.20, \delta=0.001$ & 126.0 & 97.19\% & 120 & 55.3\\
        $\epsilon=0.30, \delta=0.001$ & 131.4 & 52.43\% & 260 & 13.5\\
        $\epsilon=0.40, \delta=0.001$ & 130.1 & 94.64\% & 900 & 61.4\\
        $\epsilon=0.50, \delta=0.001$ & 124.7 & 92.81\% & 1400 & 27.0\\
        \toprule
    \end{tabular}
    \label{tab:exploration}
\end{table}

\subsubsection{\textbf{Learning Rate}} It represents the weight given to the update of the Q-values. If the step size is too small, the current Q-value estimates will likely get stuck in a local maximum. In contrast, when $\alpha$ is too large, convergence is impossible in most cases. \tablename~\ref{tab:learning} shows that adjusting the learning rate causes significant differences in the average learning time. In this sense, it was infeasible to do a complete hyperparameter exploration given the infinite value space. However, $\alpha=0.0050$ is the best configuration considering the average training time and episodes to learn the best encryption configuration (configuration 4). Although other variants achieved a higher AQD, they required more episodes and time to find configuration 4 as the best. In summary, it trades the speed against distinction capabilities as the Q-values for configuration 4 are very close to those of configurations 5 and 6. Nevertheless, given the presented selection of variants, $\alpha=0.0050$ is considered the best approximation of the optimal setting due to its clear advantage in speed and accuracy.

\begin{table}[htpb!]
    \centering
    \caption{Agent Performance for Different Learning Rates}
    \begin{tabular}{l | c c c c }
        \toprule
        \textit{Learning rate} & \textit{Learning (min)} & \textit{Accuracy} & \textit{Episodes} & \textit{AQD} \\\midrule
        $\alpha=0.0001$ & 331.3 & 00.10\% & -- & 0.7 \\
        $\alpha=0.0005$ & 124.9 & 99.24\% & 320 & 69.7 \\
        $\alpha=0.0010$ & 149.5 & 99.34\% & 200 & 74.0 \\
        \rowcolor{LightGray}$\alpha=0.0050$ & 139.4 & 99.24\% & 150 & 60.5 \\
        $\alpha=0.0100$ & 162.1 & 99.25\% & 150 & 45.7 \\
        \toprule
    \end{tabular}
    \label{tab:learning}
\end{table}

\subsubsection{\textbf{Discount Factor}} It controls the importance of future estimations over the current one. With $\gamma$ approaching 1, the agent considers future value estimations more strongly. In contrast, with $\gamma = 0$, the agent only maximizes immediate estimates. \tablename~\ref{tab:discount} shows the impact of the discount factor on the average learning time, the accuracy, the number of episodes, and the AQD values to find configuration 4 as the best one. Interestingly, the accuracy remains relatively stable for a discount factor between 0.2 and 0.6. Analyzing the results, the setting $\gamma=0.10$ is considered best as it has the shortest average training time and the third lowest number of episodes. In addition, it provides a reasonable distinction of Q-values from configuration 4 (the best) and the rest.

\begin{table}[htpb]
    \centering
    \caption{Agent Performance for Different Discount Factors}
    \begin{tabular}{l | c c c c }
        \toprule
        \textit{Discount} & \textit{Learning (min)} & \textit{Accuracy} & \textit{Episodes} & \textit{AQD} \\\midrule
        $\gamma=0.00$ & 154.1 & 54.03\% & 110 & 18.7 \\
        \rowcolor{LightGray} $\gamma=0.10$ & 139.4 & 99.24\% & 150 & 60.5 \\
        $\gamma=0.20$ & 161.8 & 99.22\% & 260 & 86.1 \\
        $\gamma=0.30$ & 158.1 & 99.29\% & 130 & 81.5 \\
        $\gamma=0.40$ & 163.2 & 99.18\% & 130 & 102.5 \\
        $\gamma=0.50$ & 162.3 & 99.29\% & 300 & 34.4 \\
        $\gamma=0.60$ & 152.0 & 99.21\% & 320 & 71.8 \\
        $\gamma=0.70$ & 156.5 & 51.80\% & 400 & 63.8 \\
        $\gamma=0.80$ & 187.0 & 51.89\% & 410 & 29.8 \\
        $\gamma=0.90$ & 207.6 & 51.79\% & 300 & 07.3 \\
        $\gamma=1.00$ & 253.0 & 49.69\% & 760 & 83.2 \\
        \toprule
    \end{tabular}
    \label{tab:discount}
\end{table}

\subsubsection{\textbf{Hidden Neurons}} \tablename~\ref{tab:layers} shows how the number of neurons in the hidden layer affects the learning time, accuracy, number of episodes, and AQD of the agent. Sizes from 10 to 50 neurons were tested, achieving the best setup configuration with 25 hidden neurons.

\begin{table}[htpb]
    \centering
    \caption{Agent Performance for Different Hidden Neurons}
    \begin{tabular}{l | c c c c }
        \toprule
        \textit{Neurons} & \textit{Learning (min)} & \textit{Accuracy} & \textit{Episodes} & \textit{AQD} \\\midrule
        10 & 155.3 & 99.46\% & 220 & 44.6 \\
        20 & 155.4 & 99.45\% & 150 & 45.2 \\
        \rowcolor{LightGray}25 & 195.8 & 99.52\% & 110 & 63.9 \\
        30 & 198.7 & 99.42\% & 240 & 51.4 \\
        35 & 200.4 & 99.33\% & 100 & 56.6 \\
        40 & 139.4 & 99.24\% & 150 & 60.5 \\
        45 & 195.5 & 99.26\% & 130 & 42.5 \\
        50 & 204.0 & 99.21\% & 580 & 53.2 \\
        \toprule
    \end{tabular}
    \label{tab:layers}
\end{table}

\subsubsection{\textbf{Activation Function}} \tablename~\ref{tab:actfunction} lists the activation functions evaluated in the first and second layers of the neural network and their impact on the learning process. As can be seen, the activation function significantly impacts the average training time. More in detail, the Log-SiLU setting obtains the best results because, although Log-ReLU achieves slightly better accuracy, many Q-values are equal to zero due to the dying ReLU problem.

\begin{table}[htpb!]
    \centering
    \caption{Agent Performance for Different Activation Functions}
    \begin{tabular}{l | c c c c }
        \toprule
        \textit{Activation Func.} & \textit{Learning (min)} & \textit{Accuracy} & \textit{Episodes} & \textit{AQD} \\\midrule
        Log - Log & 343.6 & 00.02\% & -- & -0.0022 \\
        Log - ReLU & 110.9 & 99.32\% & 350 & 49.9 \\
        \rowcolor{LightGray}Log - SiLU & 139.4 & 99.24\% & 150 & 60.5 \\
        ReLU - Log & 109.2 & 00.25\% & -- & -0.3 \\
        ReLU - ReLU & 334.3 & 00.07\% & -- & -10.5 \\
        ReLU - SiLU & 151.4 & 00.18\% & -- & -27.9 \\
        SiLU - Log & 137.2 & 00.27\% & -- & -3300 \\
        SiLU - ReLU & 361.0 & 00.11\% & -- & 0 \\
        SiLU - SiLU & 171.5 & 00.12\% & -- & -36.7 \\
        \toprule
    \end{tabular}
    \label{tab:actfunction}
\end{table}

\subsubsection{\textbf{Weights Initialization}} Plain random-uniform distribution, Xavier uniform distribution, and He initialization were tested. The plain uniform initialization randomly selects weights from a uniform distribution in the $[0,1]$ range. In the Xavier initialization, the distribution is dynamically scaled according to the dimensions of the previous layer. Lastly, He initialization generates random numbers selected from a standard normal distribution. After evaluating the three of them, Xavier provided the best learning time (105.4 min), accuracy (99.31\%), and the number of episodes (90).

\subsection{Agent Final Performance}

According to the results obtained in the previous experiment, the Agent of \solution{} is configured with the following configuration of hyperparameters: $\epsilon=0.20$, $\delta=0.01$, $\alpha=0.005$, $\gamma=0.30$, hidden neurons = 25, activation functions = \textit{Log-SiLU}. \tablename~\ref{tab:final} shows the accuracy with which the Agent selects configuration 4 for different episodes and therefore learning times. As can be seen, $>$96\% accuracy is obtained in less than 10 minutes.

\begin{table}[htpb!]
    \centering
    \caption{Agent Performance with Best Hyperparmeter Configuration}
    \begin{tabular}{l | c c }
        \toprule
        \textit{Episodes} & \textit{Learning (min)} & \textit{Accuracy} \\\midrule
        100 & 2.0 & 91.43\% \\
        200 & 5.1 & 94.86\% \\
        300 & 6.5 & 96.32\% \\
        400 & 8.1 & 96.21\% \\
        1'000 & 23.9 & 98.61\% \\
        2'000 & 66.4 & 99.07\% \\
        5'000 & 172.2 & 99.71\% \\
        \toprule
    \end{tabular}
    \label{tab:final}
\end{table}

In conclusion, the obtained results demonstrated that \solution{} is able to learn how to evade intelligent detection systems in just a few minutes and with promising accuracy. Therefore, more efforts are needed to improve detection systems against intelligent ransomware samples.
\section{Summary and Future Work}
\label{sec:conclusions}

This work proposes \solution{}, a framework able to intelligently and automatically adapt the encryption behaviors of ransomware and avoid being detected by dynamic defense mechanisms. The main contribution of \solution{} is an Agent that combines RL and device fingerprinting to learn the encryption rate, duration, and algorithm combination that maximizes encryption and minimizes detection. The learning task is driven by a reward mechanism that prioritizes the encryption rate and stealth capabilities of ransomware configurations. Stealth is evaluated using an anomaly detection system that uses ML and fingerprinting, as proposed in the literature.

\solution{} has been deployed in a real scenario composed of a Raspberry Pi 4 acting as a crowd-sensor affected by a recent ransomware called Ransomware-PoC. More in detail, the components of \solution{} have been integrated into Ransomware-PoC, which has been modified to dynamically adapt its algorithms, encryption rates, and duration. A pool of experiments combining Deep Q-Learning and Isolation Forest (in the Agent and detection system, respectively) has demonstrated that \solution{} evades the detection of Ransomware-PoC affecting the Raspberry Pi 4 in 2 minutes with $>$90\% accuracy. %In conclusion, in the testbed scenario, Ransomware-PoC is able to encrypt almost 30 MB per day without being detected. 

Future work plans to evaluate the functionality of \solution{} with different benign device behaviors to show its adaptability. It is also planned to try other malware samples, such as backdoors leaking sensitive data. Finally, it is expected to work on intelligent and adaptive detection mechanisms to detect \solution{}.

\section*{Acknowledgment}
This work has been partially supported by \textit{(a)} the Swiss Federal Office for Defense Procurement (armasuisse) with the CyberForce project (CYD-C-2020003) and \textit{(b)} the University of Zürich UZH.

\bibliographystyle{IEEEtran}  
\balance
\bibliography{references}

\end{document}